\documentclass[twocolumn,twoside,slac_two]{revtex4}
\usepackage{graphicx}
\usepackage{fancyhdr}
\newcommand{\ttbar}  {\ensuremath{t\bar{t}} }
\newcommand{\ppbar}  {\ensuremath{p\bar{p}} }
\newcommand{\qqbar}  {\ensuremath{q\bar{q}} }

\newcommand{\gevccnoarg}{\ensuremath{\mathrm{GeV}/c^{2}} }

\newcommand{\gevcc}[1]  {\ensuremath{#1~\mathrm{GeV}/c^{2}}}
\newcommand{\invpb}[1]  {\ensuremath{#1~\mathrm{pb}^{-1}}}
\newcommand{\invfb}[1]  {\ensuremath{#1~\mathrm{fb}^{-1}}}

\begin{document}

\title{Top quark properties at CDF}

%

\author{Hyun Su Lee (On Behalf of the CDF collaboration)}
\affiliation{University of Chicago, Chicago, IL 60637, USA}

\begin{abstract}
We present the top property measurements in the CDF.
Most of
measurements utilize close to the integrated luminosity of
3~fb$^{-1}$.
\end{abstract}

\maketitle

\thispagestyle{fancy}


\section{Introduction}
During the last decade after discovery of top
quark~\cite{r_run1topdiscCDF,r_run1topdiscD0},
top quark has been inclusively studied. By now, the mass of the top
quark has been measured to be
173.1$\pm$1.3~\gevccnoarg\cite{masscombo} which is the most
precisely measured quark mass and \ttbar pair production cross section
has been measured as less than 10~\% of uncertainties~\cite{cross}. However, many of
another top quark property have not yet been well explored due to the
limited statistics. In the ongoing data taking at Fermilab's Tevatron
proton-antiproton collider with Collider Detector at Fermilab~(CDF),
an increasing of integrated luminosity can make us to
measure the property and also discover the unexpected phenomena from
top sector.  We describes a few of the CDF's progress of top quark
property measurements in the following.  
\section{Top property measurements}
\subsection{Top quark production}
\begin{figure}[h]
\centering
\includegraphics[width=80mm]{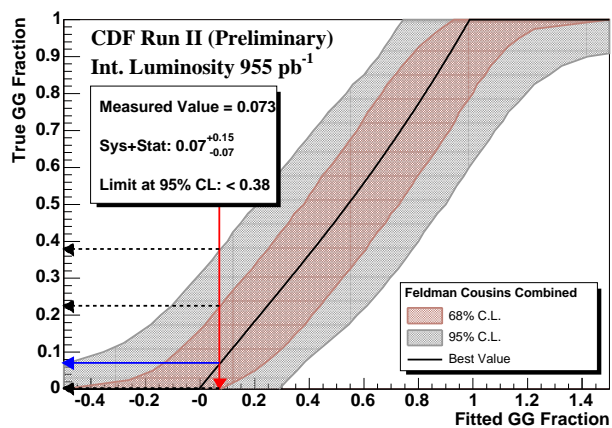}\hspace{2pc}%
\caption{\label{ggfrac_fc}
Feldman-Cousins bands for the combination of the analysis for $gg$
fraction in the lepton jet channel with
statistical and systematic uncertainties for 68\% C.L. and 95\% C.L..}
\end{figure}

The predominant production of top quark in the Tevatron is the \ttbar
pair production. 
The standard model~(SM) predicts the \ttbar production processes to be
\qqbar annihilation~($\qqbar \rightarrow \ttbar$) and $gg$ fusion~($gg
\rightarrow \ttbar$), occurring at the Tevatron with relative
fractions of $\sim$85\% and $\sim$15\%, respectively~\cite{ggfrac}. A
measurement of this fraction tests the SM predictions and our
understanding of gluon parton distribution functions in the proton. 
We measure this quantity both lepton jets and dilepton final state. In
the lepton jet channel, we have two different measurement based on
different discriminant. One method builds distriminant using number of
low-momentum track to take advantage of the higher probability for a
gluon than for a quark to radiate a low-momentum
gluon~\cite{gg_lj1}. The other
method builds discriminant with artificial neural net using eight
variables those are sensitive to the production mechanism~\cite{gg_lj2}. Both two
measurement use \invpb{955} of CDF data, and is combined by using the
Feldman-Cousin prescription~\cite{fc}. Figure~\ref{ggfrac_fc} shows
Feldman-Cousin bands for the combination of two analysis with 68\% and
95\% confidence level~(C.L.) for the fraction of $gg$ to produce
\ttbar. We measure the fraction to be $G_{f}=0.07^{+0.15}_{-0.07}$, and
we find the 95\% C.L. limit to be $G_{f} < 0.38$~\cite{gg_lj2}. This is consistent
with SM. 

\begin{figure}[h]
\centering
\includegraphics[width=80mm]{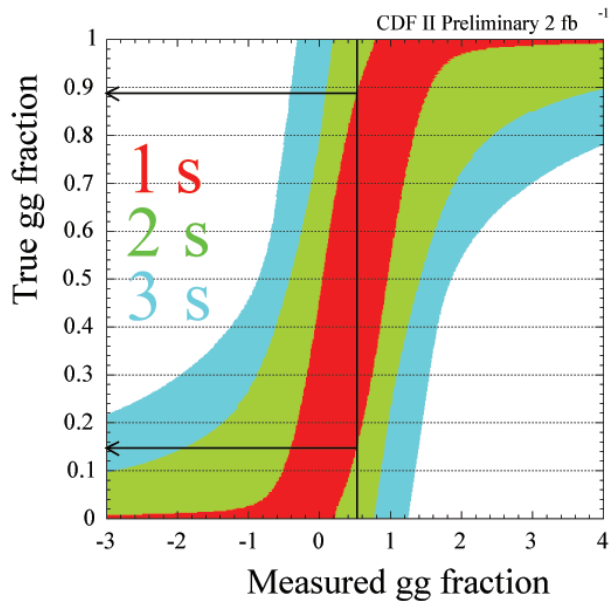}\hspace{2pc}%
\caption{\label{ggfrac_fc_dil}
Feldman-Cousins bands for $gg$ fraction in the dilepton channel with
statistical and systematic uncertainties for 68\% C.L. and 95\% C.L..}
\end{figure}

We use angle between two lepton to build discriminant for extraction
of $gg$ fraction in the dilpeton channel~\cite{gg_dil}. We build Feldman-Cousin
bands and fit data using \invfb{2.0} of \ppbar collision as you can see in
Fig.~\ref{ggfrac_fc_dil}. We measure the $gg$ fraction in the dilepton
channel to be $G_{f} = 0.53^{+0.36}_{-0.38}$ which is consistent with
SM. 

\begin{figure}[h]
\centering
\includegraphics[width=80mm]{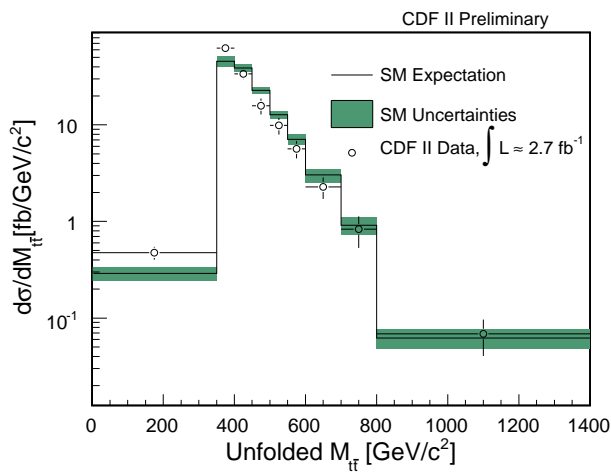}\hspace{2pc}%
\caption{\label{dsigma}
Unfolded differential cross section of \ttbar invariant mass using CDF
data is compared with SM prediction.}
\end{figure}

Since new production mechanism for top quark pairs can make the shape
of \ttbar invariant mass as resonances or general shape distortions,
the generic method to search the such contribution is to compare the
shape of the observed differential \ttbar cross
section $d\sigma/dM_{\ttbar}$ with SM expectation. The
mass of the top-antitop system is reconstructed for each event by
combining the four vectors of the four leading jets, lepton, and
missing transverse energy. The unfolding technique implemented to
correct the reconstructed distribution as for direct comparison with
theoretical differential cross section. In the update with \invfb{2.7}
data, we have in-situ jet energy scale~(JES) measurement using di-jet
mass of $W$ boson decay, which have been used in the top quark mass
measurement~\cite{mass}, that we can significantly reduce the JES systematics. As
one can see in Fig.~\ref{dsigma}, we do not find any significant difference
with SM expectation. We
check the consistency using the Anderson-Darling~(AD)
statistics~\cite{AD}. We calculate a p-value of 0.28 using AD statistics which
have a good agreement with the SM~\cite{cdfdsigma}.

\begin{figure}[h]
\centering
\includegraphics[width=80mm]{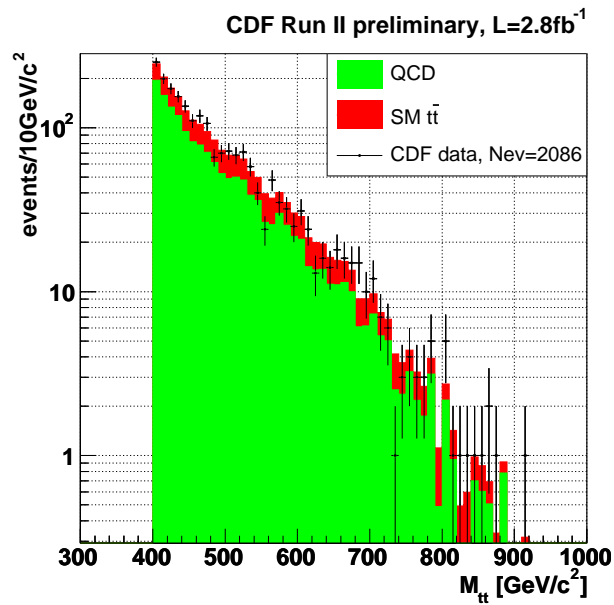}\hspace{2pc}%
\caption{\label{mtt}
Invariant mass of \ttbar from data and SM expectation.}
\end{figure}

We search for resonant top-antitop pair production and subsequent
decay in the all-hardronic channel with \invfb{2.8} of
data~\cite{mttbar}. We use
the probability from per-event matrix element calculation as
discriminant to reduce and control the large background from
QCD-multijet as an input of neural net with other kinematic variables.
We reconstruct invariant mass of \ttbar with matrix element technique
and have consistent result with SM prediction as one can see in
Fig.~\ref{mtt}. We then set the 95\% C.L. limits on Z' production as
805~GeV in case of a leptophobic topcolor resonance candidate. 

\subsection{Top like new physics particle search}
\begin{figure}[h]
\centering
\includegraphics[width=80mm]{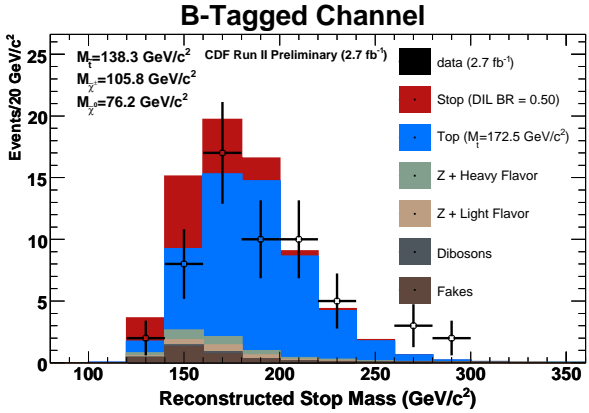}\hspace{2pc}%
\caption{\label{stop}
Reconstructed stop mass comparing data to monte carlo.}
\end{figure}

Due to the large mass of the top quark, the super-symmetric partner of
top quark~(stop) can be lighter than the top quark even to be a
lightest squark. For a light stop and R-parity conservation of
super-symmetry particles, stop quark dominantly decay to b-quark and
chargino, and chargino decay to W boson and neutralino. While
neutralino can not be detected, the decay of pair produced stop quark
have same final state with \ttbar decay. 
We search the pair production of stop in the dilepton final state with
\invfb{2.7} of data~\cite{stops}.
We reconstruct the stop mass in the
underconstraint system to extract the stop components in the \ttbar
dilepton decay events. Figure~\ref{stop} shows our data is compatible
with SM prediction without stop, and then we set the 95\% C.L.
upper limit of stop in certain condition of SUSY parameter space~\cite{stops}. 

\subsection{Top properties}
\begin{figure}[h]
\centering
\includegraphics[width=80mm]{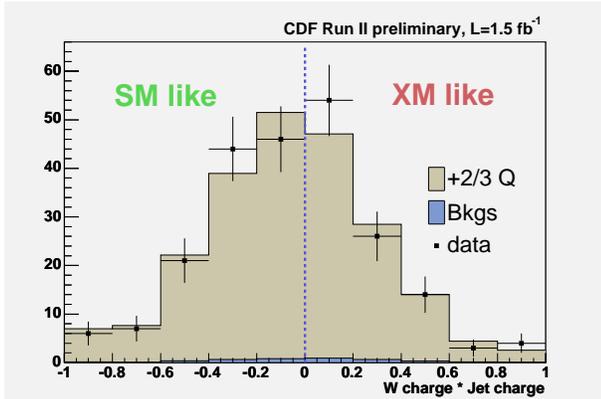}\hspace{2pc}%
\caption{\label{charge}
Product of the W charge and the associated jet charge for data and
MC~(SM signal MC distribution).}
\end{figure}

One of basic quantities of top quark is the electric charge, which is
expected to have a value of 2/3$e$ in the SM. However, one of exotic
model have decay of top to a $W^{-}$ instead of $W^{+}$ having a
charge of 4/3$e$~\cite{tcex}. We have a measurement using \invfb{1.5} of data in
the lepton jet channel. The measurement identifies the charge of the
two W bosons and two b-quarks in each data event, and then determine
which W bosons
and b-quarks decayed from the same parent top quark. The charge of the
top is then obtained by multiplying the charge of the W with the
charge of the jet associated with a b-quark. Figure~\ref{charge} shows
the measured distribution of pairs of charge product which is
compatible with SM like top charge. We then exclude exotic model-like
top at 87\% C.L.~\cite{tcharge}.

\begin{figure}[h]
\centering
\includegraphics[width=60mm]{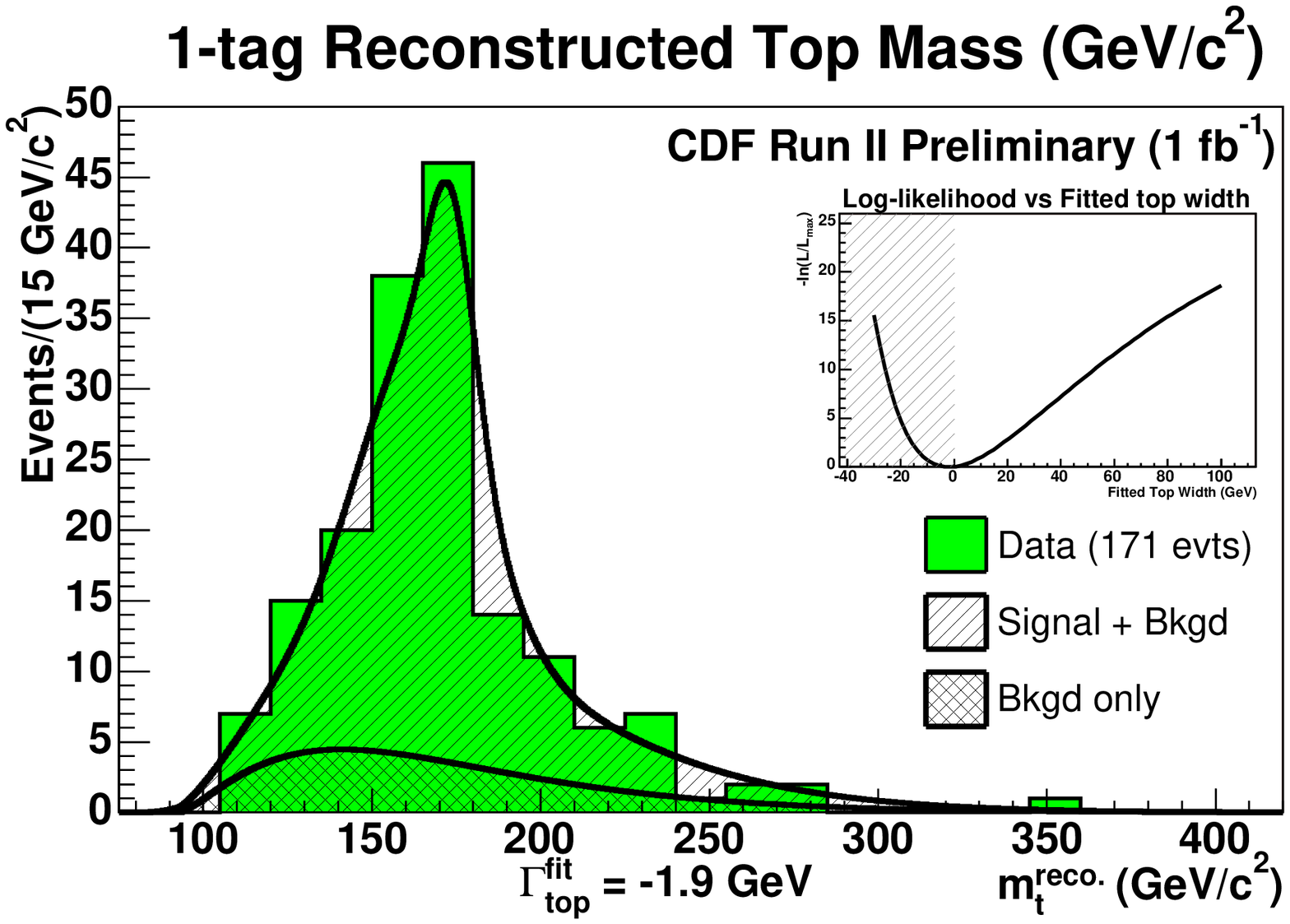}\hspace{2pc}%
\includegraphics[width=60mm]{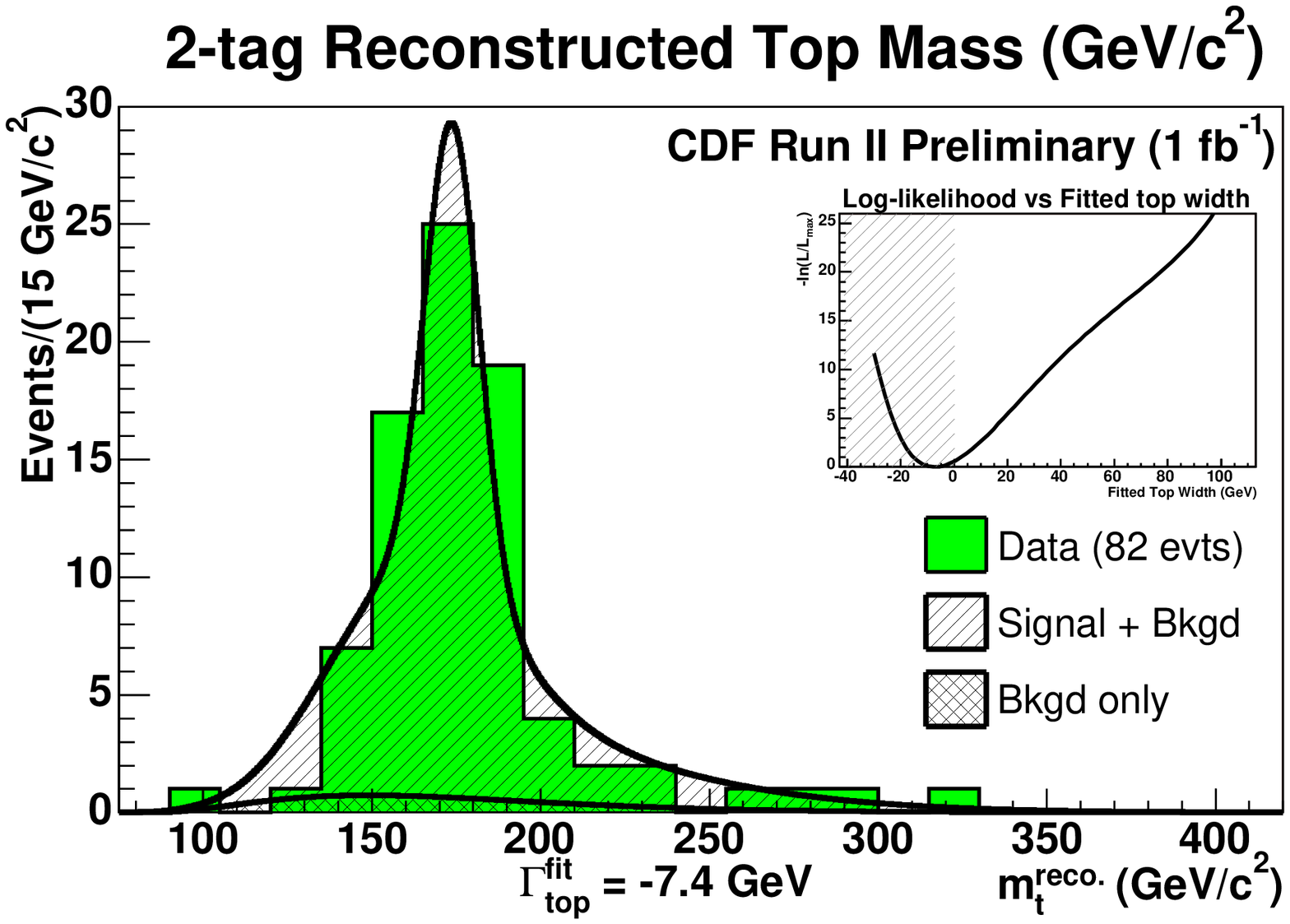}\hspace{2pc}%
\caption{\label{width}
Likelihood fit to the data with 1-tag~(up) and 2-tag~(down) categories.}
\end{figure}

Top quark width have been measured with \invfb{1} of data in the
lepton jet channel. Main idea is to use top mass reconstruction and
templates for different top width and to fit it to data. We use the
top quark mass considered known as $M_{top} = \gevcc{175}$ and
templates are produced for range of different top width. We extract
top width from reconstructed top mass distribution compared to signal
with different top width and background using unbinned likelihood fit.
We then have measurement consistent with SM as one can see in
Fig.~\ref{width}, so we set a limit on top width using
Feldman-Counsins~\cite{fc} prescription to be top width($\Gamma$)$<$13.1~GeV of
95\% C.L. upper limit~\cite{twidth}. 

\subsection{Top decay}
\begin{figure}[h]
\centering
\includegraphics[width=60mm]{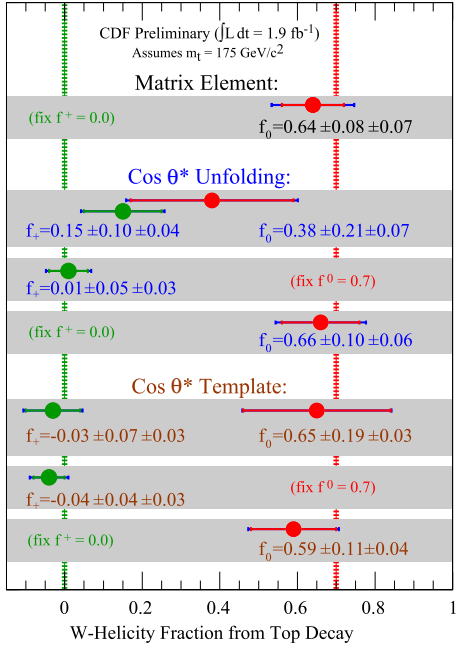}\hspace{2pc}%
\caption{\label{whel}
The summary of W-helicity measurement at CDF.}
\end{figure}

The SM predicts that the top quark decays almost entirely to a W-boson
and a bottom quark, and that the Wtb vertex is a V-A charged weak
current interaction. A consequence of this is that the top quark is
expected to decay 70.4\% of the time of longitudinal and the rest to
left handed polarized W-bosons~\cite{wpred}. Any new particles involved in the same
decay topologies and non-standard coupling could create a different
mixture of polarized W-bosons. Therefore, a measurement of this
fraction is a test of the V-A nature of the Wtb vertex. In the CDF,
there are several measurements using different technique in the lepton
jet channel with \invfb{1.9} of data. One method builds template using
$cos(\theta)$~\cite{wtemp}, where the
$\theta$ is the angle between lepton and b-quark in the W rest frame,
this is sensitive to W helicity.
The other method use matrix element technique~\cite{wmat} which we calculate a
likelihood for each event then product per event likelihood to build
total likelihood. Figure~\ref{whel} shows W-helicity measurements at CDF.
All of measurements are consistent with SM prediction. 

\begin{figure}[h]
\centering
\includegraphics[width=80mm]{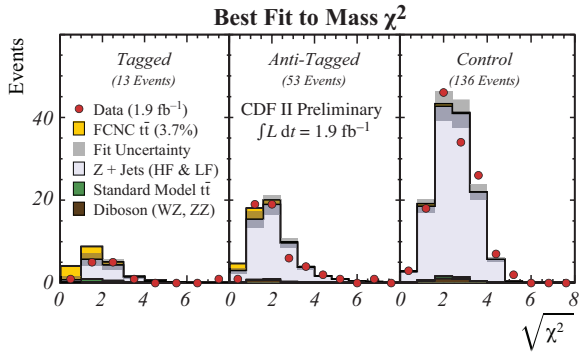}\hspace{2pc}%
\caption{\label{fcnc}
Mass $\chi^2$ distribution for signal and control regions.}
\end{figure}

Several exotic physics models, such as SUSY and two Higgs doublet,
predict flavor changing neutral currents~(FCNC) in top decay. In the
standard model, this decay mode is highly suppressed. Therefore, any
signal from FCNC decay chain indicate an evidence of new physics. A
search for FCNC decays has been performed at CDF with \invfb{1.9}. This
analysis utilizes a template fit to a mass $\chi^2$ variable
constructed from kinematic constraints present in FCNC top quark
decays. A simultaneous fit is performed to the data using two signal
and one control region as one can see in Fig.~\ref{fcnc}. The control region constrains uncertainties in
the shape and normalization of the templates. As one can see in
this plot, our data is well explained without FCNC components and
then, we set 95\% C.L. upper limit on the branching fraction of
($t\rightarrow Zq$) $<$ 3.7\%~\cite{tfcnc}.

\begin{figure}[h]
\centering
\includegraphics[width=60mm]{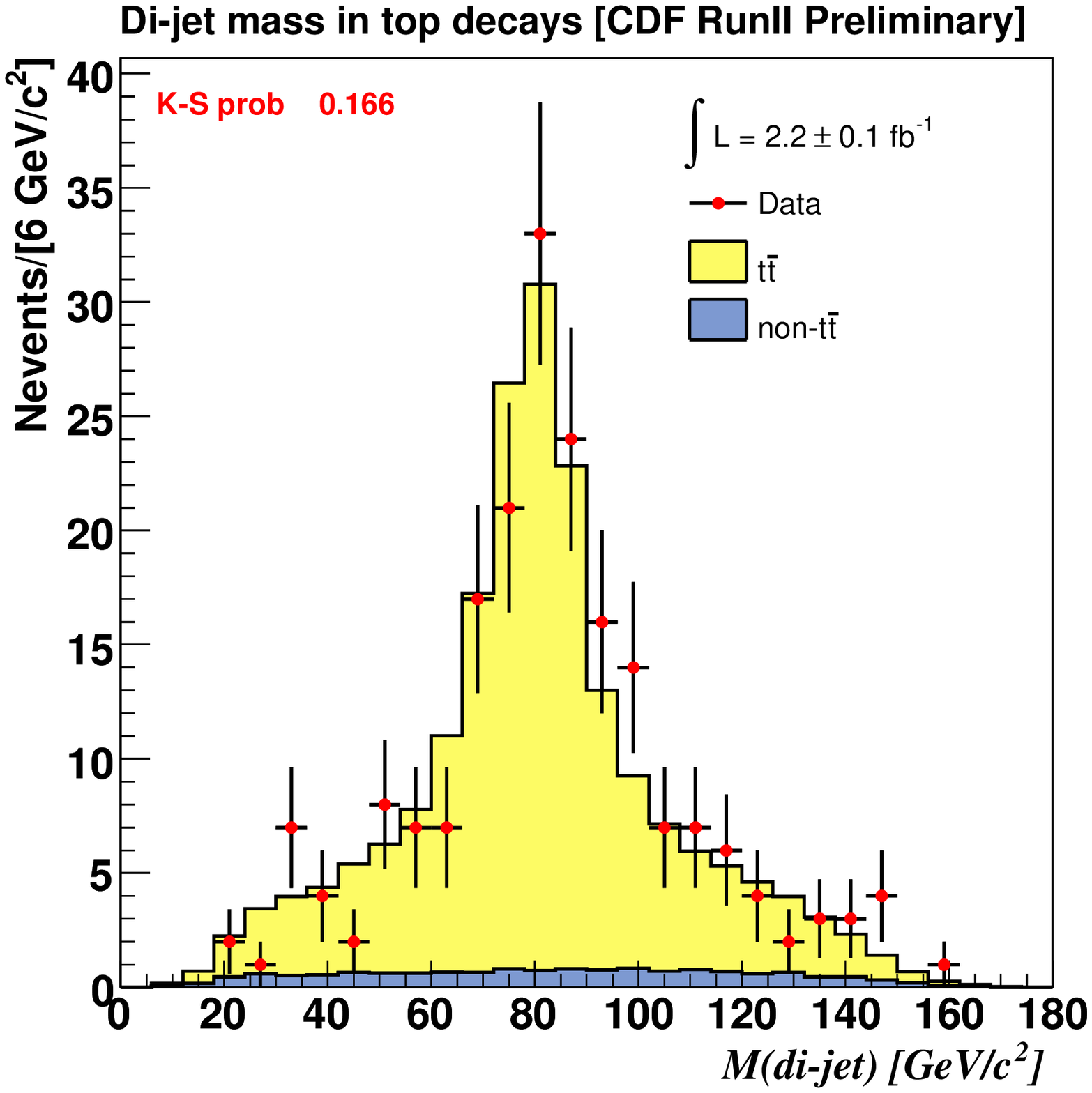}\hspace{2pc}
\includegraphics[width=60mm]{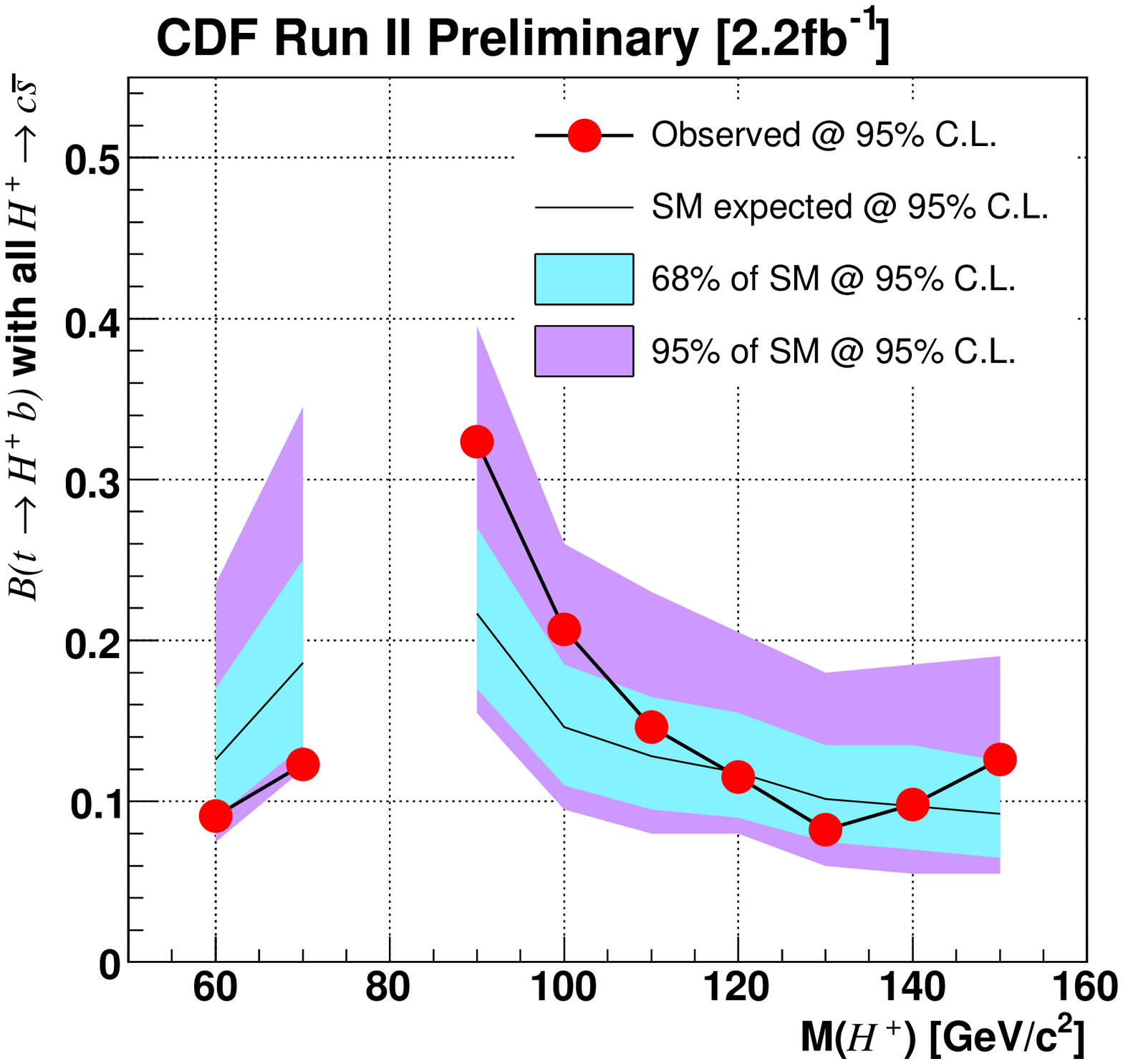}\hspace{2pc}
\caption{\label{chiggs}
Distribution of di-jet mass in \ttbar decay~(up) and the upper limit
on the branching fraction of $t\rightarrow H^{+}b$ at 95\% C.L. as a
function of charged higgs mass~(down).}
\end{figure}

Charged Higgs $H^{\pm}$ bosons are predicted in supersymmetric and GUT
extensions of the SM. If a charged Higgs boson in sufficiently light,
it can be produced in top quark decays. In the presence of a charged
Higgs boson, the $t \rightarrow H^{+}b$ decay would compete with the
SM top quark decay, thereby altering the expected number of events in
different final states of \ttbar. In the certain final state, which is
low $\tan \beta$, the dominant decay of charged higgs is
$H^{+}\rightarrow c\bar{s}$. We has searched for the decays in the
lepton + jets events with \invfb{2.2} by fully reconstructing \ttbar decay and
exploiting the difference between the dijet mass spectra in $W
\rightarrow q\bar{q}$ and $H^{+} \rightarrow c\bar{s}$
decays~\cite{chigg}. The
invariant dijet mass spectrum in data is shown in
Fig.~\ref{chiggs}~(up)
which no significant deviation from the SM is observed. Therefore we
set the limits on branching fraction of $t\rightarrow
H^{+}b\rightarrow c\bar{s}b$ as one can see in
Fig.~\ref{chiggs}~(down). In this plot, we extend our search to
generic charged boson search which possibly have smaller mass than W
boson.  

\section{Conclusions}
Number of top quark properties not only standard model top signature
but also exotic model signature
have been searched and measured. However many measurements are still
limited by the statistical uncertainty. Although we do not find
evidence conflicting with SM top
quark, we expect to have interesting
measurement with more data in near future. 

\begin{acknowledgments}
I would like to thank the CDF colleagues for their effort to carry out
these challenging physics analysis. I also thank the conference
organizers for a very rich week of physics. 
\end{acknowledgments}

\bigskip 

\end{document}